\documentclass[12pt]{article}
\usepackage{amsmath}
\usepackage[margin=1in]{geometry}
\usepackage{amssymb}
\usepackage{amsfonts}
\usepackage{mathrsfs}
\usepackage{amsthm}
\usepackage{tikz-cd}
\begin{document}

\title{Determinism and Asymmetry in General Relativity}
\author{JB Manchak, TW Barrett, HP Halvorson, JO Weatherall}

\date{}

\maketitle

\hrule
\vspace{.3cm}
\noindent This paper concerns the question of which collections of general relativistic spacetimes are deterministic relative to which definitions. We begin by considering a series of three definitions of increasing strength due to Belot. The strongest of these definitions is particularly interesting for spacetime theories because it involves an asymmetry condition called `rigidity' that has been studied previously in a different context by Geroch. We show how Belot's strongest definition is connected to a formal result emphasized by Weatherall in his paper on the hole argument. We then go on to explore other (stronger) asymmetry conditions that give rise to other (stronger) forms of determinism. We introduce a number of definitions of this type and clarify the relationships between them and the three considered by Belot. We show that there are collections of general relativistic spacetimes that satisfy much stronger forms of determinism than previously known. We also highlight a number of open questions.
\vspace{-.3cm}

\hrule

\vspace{.5cm}

\begin{center}
\large \textbf{1. Introduction}
\end{center}

\vspace{.3cm}

\noindent This paper concerns the question of which collections of general relativistic spacetimes are deterministic relative to which definitions. Consider an intuitive idea developed in various ways by Montague ([1974), Lewis ([1983]), and Earman ([1986]): a collection of possible worlds is `MLE deterministic' if, for any worlds in the collection that agree on some initial segment, the worlds agree entirely. Belot ([1995]) considers three natural ways of making MLE determinism precise. Within the discussions of the hole argument (Earman and Norton [1987]), the first two definitions have long played a central role (Butterfield [1989]; Melia [1999]; Pooley [2021]). Belot's third definition is much stronger than the other two and has only recently become a focus of attention (Landsman [2023]; Cudek [2023], Halvorson and Manchak [2025]). For each definition, Belot provides a highly symmetric example that seems to be (i) deterministic according to the particular formulation but (ii)  indeterministic by the lights of certain haecceitist positions. This leads him to conclude that all three definitions are inadequate. 

In a recent paper, Halvorson et al. ([unpublished]) push back on this position. They hold that Belot's third definition is actually a promising precisification of MLE determinism that was unfortunately ignored for thirty years. They also argue that Belot's third example is ambiguous; until it is made precise, it is neither deterministic nor indeterministic according to his third definition. It is emphasized that whether or not a theory is deterministic depends crucially on the precise formulation of the theory and the precise formulation of determinism. 

Here in this paper, we follow this line but focus our attention on (variants of) general relativity. Belot's third definition is particularly interesting for spacetime theories because it involves an asymmetry condition called `rigidity' that has been studied previously in a different context (Geroch [1969]; Halvorson and Manchak [2025]; Dewar [2025]). We show how this definition is connected to a formalresult emphasized by Weatherall ([2018]) in his paper on the hole argument. We then explore other (stronger) asymmetry conditions that give rise to other (stronger) forms of MLE determinism. We introduce a number of definitions of this type and clarify the relationships between them and the three considered by Belot. We go on to show that there are collections of general relativistic spacetimes that satisfy much stronger forms of determinism than previously known. We also highlight a number of open questions.

\vspace{.3cm}

\begin{center}
\large \textbf{2. De Dicto and De Re Determinism}
\end{center}

\vspace{.3cm}

\noindent Let $\mathscr{U}$ be the collection of all general relativistic spacetimes $(M, g)$. Here $M$ is a smooth, connected, Hausdorff, $n$-dimensional (for $n\geq 2$) manifold and $g$ is a smooth Lorentzian metric on $M$. In order to make sense of the notion of `initial segment' in the MLE definition of determinism, we first restrict attention to the collection $\mathscr{H} \subset \mathscr{U}$ of globally hyperbolic spacetimes with a chosen temporal orientation (see Wald [1984]). Any $n$-dimensional globally hyperbolic spacetime $(M,g)$ admits an $(n-1)$-dimensional Cauchy surface $\Sigma \subset M$ representing all of space at a given time. One can show that $M$ can be foliated by a collection of Cauchy surfaces such that the topology of $M$ is $\Sigma \times \mathbb{R}$. For any spacetime $(M,g) \in \mathscr{H}$, we define an initial segment $U \subset M$ to be the timelike past $I^-[\Sigma]$ of any Cauchy surface $\Sigma$ in $(M,g)$. One can verify that any initial segment $U$ of any globally hyperbolic spacetime $(M,g)$ counts as a connected, open subset of $M$. So $(U,g)$ is a spacetime in its own right. Let us now consider two forms of MLE determinism adapted from Belot ([1995]):  \\

\noindent {\bf Definition 1:} A collection $\mathscr{C} \subseteq \mathscr{H}$ is de dicto deterministic if, for any spacetimes $(M,g)$, $(M',g') \in \mathscr{C}$ and any initial segments $U \subset M$ and $U' \subset M'$, if there is an isometry $\varphi: U \rightarrow U'$, then there is an isometry $\psi: M \rightarrow M'$. \\

\noindent {\bf Definition 2:} A collection $\mathscr{C} \subseteq \mathscr{H}$ is de re deterministic if, for any spacetimes $(M,g)$, $(M',g') \in \mathscr{C}$ and any initial segments $U \subset M$ and $U' \subset M'$, if there is an isometry $\varphi: U \rightarrow U'$, then there is an isometry $\psi: M \rightarrow M'$ such that $\psi_{|U}=\varphi$. \\

These two forms of determinism (or close variations) go by different names and have long played a central role in discussions of the hole argument. The first is called `Dm2' in (Butterfield [1989]), `Definition 1' in (Belot [1995]), `Lewis's analysis' in (Melia [1999]), `Det2' in (Pooley [2021]), and `D1' in (Halvorson et al. [unpublished]). The second is called `Definition 2' in (Belot [1995]), the `second resolution' in (Melia [1999]), `Dm2+' in (Cudek [2023]), and `D2' in (Halvorson et al. [unpublished]). Here, we use the names `de dicto' and  `de re' from (Dewar [2016], [2025]). The two forms of determinism capture slightly different senses in which a collection of spacetimes might be considered deterministic. Clearly, de re determinism implies de dicto determinism. They differ in how one captures `agreement of worlds' in the MLE formulation of determinism. To better understand the definitions, consider the following example. \\

\noindent {\bf Example 1:} Consider two-dimensional Minkowski spacetime in standard $(t,x)$ coordinates and let $(M, g)$ be the $t<0$ region (cf. Roberts [2020]; Manchak [2020]). The collection $\mathscr{C}=\{(M,g)\}$ is de dicto deterministic. This follows easily since there is only one spacetime in the collection. But $\mathscr{C}$ is not de re deterministic. To see this, consider the initial segments $U, U' \subset M$ given by the $t<-2$ and $t<-1$ regions respectively. Clearly there is a time translation isometry $\varphi: U \rightarrow U'$ defined by $\varphi(t,x)=(t+1, x)$. But there can be no corresponding global isometry $\psi: M \rightarrow M$ such that $\psi_{|U}=\varphi$ because of the $t=0$ `edge' of the spacetime (see Figure 1). \\   

\begin{figure}[htp!] \centering
 \includegraphics[width=4in]{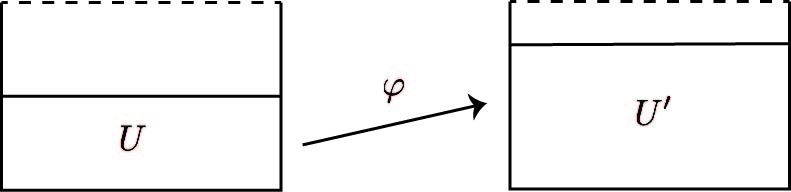} 
 \caption{Two copies of the spacetime $(M,g)$ from Example 1. The time translation isometry $\varphi$ from the initial segment $U$ to the initial segment $U'$ does not extend to a global isometry because of the $t=0$ `edge' of the spacetime (dotted line).}.
\end{figure}

From Example 1, we have the following: \\

\noindent {\bf Proposition 1:} If a collection $\mathscr{C} \subseteq \mathscr{H}$ is de re deterministic, then it is de dicto deterministic; the other direction does not hold.\\

Example 1 illustrates a curious state of affairs. The agreement of the initial segments $U$ and $U'$ given by $\varphi$ does not extend to a global agreement. So one form of MLE determinism (de re) does not hold. And yet there is global agreement given by the identity map on $M$ showing that another form of MLE determinism (de dicto) does hold. A similar situation  arises in the  `collapsing columns' example due to Wilson ([1993]) often used to distinguish the two definitions. Here is Belot ([1995], p. 191):

\begin{quotation}
But now consider a world $W$, in which centrally loaded columns collapse
by buckling, which contains nothing but: (i) a large, homogeneous,
perfectly spherical planet; (ii) a relatively small perfectly cylindrical
column resting on the planet so that its axis of symmetry is normal to
the planet at the point of contact; (iii) and a cone, which is moving through
space in such a way that at time $t = 0$ its apex will impact the column at the
exact center of its top surface, with sufficient force to cause the column to
buckle. We also assume that W has a Newtonian spacetime structure; that
the laws concerning motion are completely deterministic; and that the law
governing column collapse fully determines the shape that the column
assumes upon collapse.
\end{quotation}

The classical (i.e. non-relativistic) analogue to de dicto determinism is satisfied by the singleton collection consisting of the collapsing columns world $W$. However, by the lights of certain haecceitist positions, there is indeterminism present in the example: because of the symmetries of the $t<0$ initial segment, the columns `could have toppled in a different direction' (Melia [1999], p. 649). Accordingly, the de dicto form of MLE determinism is considered not strong enough. With respect to the classical analogue of the de re definition however, the collapsing columns example fails to be de re deterministic. To see this, note that because of the symmetries of the $t<0$ portion of $W$, there will be a rotational isomorphism of the model (that preserves both the metric and matter content) from this initial segment to itself that takes any spatial direction into any other. But unless this isomorphism is the identity map, it cannot be extended to all of $W$ since the direction of collapsed column after $t=0$ must be taken into itself. We note that this example can be considered a violation of Curie's principle which states that `when certain causes produce certain effects, the elements of symmetry of the causes must be found in the produced effects' (Curie [1894], p. 401). For more on Curie's principle and general relativity, see Earman ([2007]), Roberts ([2016]), and Manchak ([unpublished]). 

Which collections of $\mathscr{C} \subseteq \mathscr{H}$ satisfy de re determinism? Consider the collection $\mathscr{V}^+ \subset \mathscr{H}$ of four-dimensional, inextendible, globally hyperbolic, vacuum solutions to Einstein's equation. Let $(M,g)$ and  $(M',g')$ be any spacetimes in $\mathscr{V}^+$ and let $U \subset M$ and $U' \subset M'$ be any initial segments. If there is an isometry $\varphi: U \rightarrow U'$, it follows from a theorem by Choquet-Bruhat and Geroch ([1969]) that there is an isometry $\psi: M \rightarrow M'$ such that $\psi_{|U}=\varphi$. Thus, we have the following which captures a sense in which significant swathes of general relativistic spacetimes of physical interest are deterministic. (We note here that the result can be generalized to include collections of non-vacuum spacetimes as well. The interested reader can consult Wald ([1984], pp. 266-267) for details.) \\

\noindent {\bf Proposition 2:} Any collection $\mathscr{C} \subseteq \mathscr{V}^+$ is de re (and thus de dicto) deterministic.

\vspace{.3cm}

\begin{center}
\large \textbf{3. De Re + Uniqueness}
\end{center}

\vspace{.3cm}

\noindent Given a collection of $\mathscr{C} \subseteq \mathscr{H}$, both the de dicto and de re forms of MLE determinism require that for any spacetimes $(M,g), (M',g') \in \mathscr{C}$ and any initial segments $U \subset M$ and $U' \subset M'$, if there is an isometry $\varphi: U \rightarrow U'$, then a certain type of isometry $\psi: M \rightarrow M'$ must exist. In the natural way, one can strengthen each definition by adding a uniqueness clause to $\psi$. Indeed, the third form of MLE determinism considered by Belot ([1995]) is de re determinism with uniqueness: worlds must agree in only one way. Consider the following:  \\

\noindent {\bf Definition 3.} A collection $\mathscr{C} \subseteq \mathscr{H}$ is de re* deterministic if, for any spacetimes $(M,g)$, $(M',g') \in \mathscr{C}$ and any initial segments $U \subset M$ and $U' \subset M'$, if there is an isometry $\varphi: U \rightarrow U'$, then there is a unique isometry $\psi: M \rightarrow M'$ such that $\psi_{|U}=\varphi$. \\

Curiously, this definition has received relatively little attention in the hole argument literature until just recently. De re* determinism is called  `Definition 3' in (Belot [1995]) and `Dm2++' in (Cudek [2023]). It is endorsed by Halvorson et al. ([unpublished]) who call it `D3' determinism. The definition is also briefly considered by Halvorson and Manchak ([2025]) and Landsman ([2023]). 

Why might one be interested in de re* determinism? It has been argued that de re determinism is not strong enough to count all instances of indeterminism as such. Belot ([1995]) introduces his own symmetric example which leads to the formulation of his Definition 3 (de re* determinism). We will return to a similar example in due course. For now let us turn to another case: classical Leibnizian spacetime (Earman [1989]). One can verify that the spacetime (considered as a singleton collection) counts as deterministic under the classical analogue of the de re formulation. And yet there are reasons to think that such a spacetime ought to be considered indeterministic. Here is Earman ([1977], p. 96):

\begin{quotation}
If the history of a particle is represented by a timelike world line (i.e. a world line which is everywhere oblique to the planes of simultaneity) on [Leibnizian spacetime], then determinism cannot hold. For among the automorphisms [of Leibnizian space-time] are those which are the identity on the portion of [Leibnizian spacetime] on or below some given time slice but which differ from the identity above; such a mapping leaves fixed the entire past history of the particles while changing their future behavior. Since the automorphisms of the space-time should be symmetries of the dynamical laws (whatever they are), there will be two solutions which describe the same particle histories for all past times but which describe different future behaviors.
\end{quotation}

Whether or not Leibnizian spacetime (or theories of matter set therein) ought to count as deterministic in some sense is an interesting question (see Stein [1977]; Weatherall [2020]). We do not enter the debate here. At this stage, we emphasize that we are not interested in capturing, once and for all, a single notion of `determinism' within spacetime theories. There are many interesting forms of determinism to explore (Doboszewski [2019]; Smeenk and W\'uthrich [2021]). In a pluralistic way, we are simply interested in mapping out which collections of general relativistic spacetimes are deterministic relative to which definitions. Here, we draw attention to the fact that there are definitions of determinism -- in particular de re* -- which are strong enough to capture Earman's sense in which Leibnizian spacetime is indeterministic. 

Let us now return to general relativity. The isometry uniqueness clause which is added to de re determinism to obtain de re* determinism has the effect of requiring a certain type of spacetime asymmetry known as `rigidity' (Geroch [1969]) which has recently become a focus of attention in the hole argument literature (Menon and Read [2023]; Dewar [2025]; Halvorson and Manchak [2025]). Consider the following definition: \\

\noindent {\bf Definition 4:} A collection $\mathscr{C} \subseteq \mathscr{U}$ is rigid if, for any $(M,g), (M',g') \in \mathscr{C}$ and any open set $O \subset M$, if there are isometries $\varphi, \psi: M \rightarrow M'$ such that $\varphi_{|O}=\psi_{|O}$, then $\varphi=\psi$. \\

If a collection of spacetimes is rigid, then for any pair of spacetimes in the collection, if there are isometries from one to the other that agree on an open set, they must agree entirely. It is easily verified that any collection $\mathscr{C} \subseteq \mathscr{H}$ that is both rigid and de re deterministic must be de re* deterministic. Indeed, one can think of the rigidity condition as a type of `dual' to de re determinism: the latter requires the existence of a certain isometry $\psi: M \rightarrow M'$ while the former requires (because initial segments are open sets) the uniqueness of such an isometry (Dewar [2025]). One can use a general rigidity result due to Geroch ([1969], Theorem A1) to show the following (Halvorson and Manchak [2025], Theorem 1):\\

\noindent {\bf Proposition 3:} Any collection $\mathscr{C} \subseteq \mathscr{U}$ is rigid. \\

One way to understand the significance of the proposition is as follows: within the context of general relativity, there can be no collection of spacetimes that is de re deterministic but not de re* deterministic (Cudek [2023], footnote 46). To distinguish between these two notions of determinism, one has to look to certain classical spacetime theories (e.g. the Leibnizian case mentioned above) or non-standard general relativity such as its non-Hausdorff generalization (Manchak and Barrett [forthcoming]). From Proposition 3, we have the following: \\

\noindent {\bf Corollary 1:} A collection $\mathscr{C} \subseteq \mathscr{H}$ is de re deterministic if and only if it is de re* deterministic. \\

Recall that $\mathscr{V}^+ \subset \mathscr{H}$ is the collection of four-dimensional, inextendible, globally hyperbolic, vacuum solutions of Einstein's equation. From Corollary 1 and Proposition 2, we have the following result which shows that significant swathes of general relativistic spacetimes of physical interest satisfy a much stronger form of determinism than previously known (Landsman [2022], footnote 23): \\

\noindent {\bf Corollary 2:} Any collection $\mathscr{C} \subseteq \mathscr{V}^+$ is de re* deterministic.\\

We close by highlighting how de re* determinism is related to a point emphasized by Weatherall ([2018], p. 336) in his paper on the hole argument. There, a formal proposition within the context of (standard) general relativity is central: if $(M,g)$ is a spacetime and $\varphi: M \rightarrow M$ is a non-trivial hole diffeomorphism, then the identity map on $M$ is not an isometry from $(M,g)$ to $(M, \varphi_*g)$. Let us articulate a general version of this statement with respect to any spacetime theory $\mathscr{T}$ for which one can make sense of initial segments. Such a theory will be a collection of models of the form $(M, O_1,...,O_n)$ where $M$ is a manifold and $O_1,...,O_n$ are geometric objects on $M$. Consider the following condition on $\mathscr{T}$. \\

\noindent {\bf (Id $\neq$ Iso):} Let $(M, O_1,...,O_n)$ be model in $\mathscr{T}$ and let $U \subset M$ be an initial segment. If $\varphi: M \rightarrow M$ is a non-trivial hole diffeomorphism such that $\varphi_{|U}$ is the identity map on $U$, then the identity map on $M$ is not an isomorphism from $(M, O_1,...,O_n)$ to $(M, \varphi_*O_1,...,\varphi_*O_n)$. \\

How is (Id $\neq$ Iso) related to de re* determinism? If the latter is generalized in the natural way to apply to any spacetime theory $\mathscr{T}$ with initial segments, then a simple result follows: if  $\mathscr{T}$ satisfies de re* determinism, then $\mathscr{T}$ must also satisfy (Id $\neq$ Iso). To see this, suppose (Id $\neq$ Iso) fails for some spacetime theory $\mathscr{T}$. So there is a model $(M, O_1,...,O_n)$ in $\mathscr{T}$, an initial segment $U \subset M$, and a non-trivial hole diffeomorphism $\varphi: M \rightarrow M$ such that $\varphi_{|U}$ is the identity map on $U$ and yet the identity map $\psi$ on $M$ is an isomorphism from $(M, O_1,...,O_n)$ to $(M, \varphi_*O_1,...,\varphi_*O_n)$. We know that $\varphi\neq \psi$ since $\varphi$ is non-trivial. Since the distinct maps $\varphi$ and $\psi$ are both isomorphisms from $(M, O_1,...,O_n)$ to $(M, \varphi_*O_1,...,\varphi_*O_n)$ and $\varphi_{|U}=\psi{|U}$, it follows that $\mathscr{T}$ fails the uniqueness clause in the de re* form of determinism. 

Note that a similar result does not hold for (a generalized form of) de re determinism. Because this definition lacks the uniqueness clause, it is not strong enough to ensure that (Id $\neq$ Iso) is satisfied. To see this, just consider Leibnizian spacetime (taken as a singleton collection). As mentioned above, this spacetime satisfies the classical analogue of de re determinism. But since Leibnizian spacetime is not rigid, one can easily verify that it fails to satisfy (Id $\neq$ Iso). 

\vspace{.3cm}

\begin{center}
\large \textbf{4. De Dicto + Uniqueness}
\end{center}

\vspace{.3cm}

\noindent As we have seen, Belot ([1995]) naturally strengthens de re determinism (his Definition 2) by introducing a uniqueness clause to obtain de re* determinism (his Definition 3). In the analogous way, he could have also considered a natural strengthening of the de dicto form of determinism (his Definition 1). Consider the following: \\

\noindent {\bf Definition 5:} A collection $\mathscr{C} \subseteq \mathscr{H}$ is de dicto* deterministic if, for any spacetimes $(M,g)$, $(M',g') \in \mathscr{C}$ and any initial segments $U \subset M$ and $U' \subset M'$, if there is an isometry $\varphi: U \rightarrow U'$, then there is a unique isometry $\psi: M \rightarrow M'$. \\

We have seen that within the context of (standard) general relativity, de re determinism and de re* determinism are equivalent. This is not the case for the de dicto and de dicto* types of determinism; we will soon show that the latter condition is strictly stronger than the former. First, we establish that the de re* and de dicto* forms of MLE determinism are independent conditions in the sense that neither implies the other. Consider the following examples: \\

 \noindent {\bf Example 2:} Let $(M,g)$ be four-dimensional Minkowski spacetime. The collection $\mathscr{C}=\{(M,g)\}$ is de re* deterministic as a consequence of Corollary 2. So the collection is de re deterministic from Corollary 1 and hence de dicto deterministic from Proposition 1. But $\mathscr{C}$ cannot be de dicto* deterministic since there are non-trivial isometries $\psi: M \rightarrow M$ of Minkowski spacetime to itself (e.g. rotations, translations). \\
 
 \noindent {\bf Example 3:} Let $(M,g)$ be two-dimensional Minkowski spacetime in standard $(t,x)$ coordinates. Let $M_1$ be the $t<0$ region of $M$ and let $M_2$ be the $t<x$ region of $M$. Consider the spacetime $(M', g)$ where $M'=M_1 \cup M_2$ (see Figure 2). One can verify that $(M', g)$ is a globally hyperbolic spacetime and, due to the `missing' region, the only  isometry $\psi: M' \rightarrow M'$ is the identity map. It follows that the collection $\mathscr{C}=\{(M',g)\}$ must be de dicto* deterministic. But the collection fails to be de re deterministic. To see this, consider the initial segment $U \subset M'$ given by the $t<-1$ region. Clearly there is a reflection isometry $\varphi: U \rightarrow U$ defined by $\varphi(t,x)=(t, -x)$. But since the only isometry $\psi: M' \rightarrow M'$ is the identity map, we find that $\psi_{|U}\neq \varphi$. Since de re determinism is not satisfied, de re* determinism is also not satisfied. (Similar examples showing the satisfaction of de dicto* determinism while violating de re determinism can be constructed even if one were to restrict attention to `maximal' spacetimes. See Manchak ([2016], [2017], [2023]) for more on this spacetime property.)  \\

 \begin{figure}[htp!] \centering
 \includegraphics[width=1.5in]{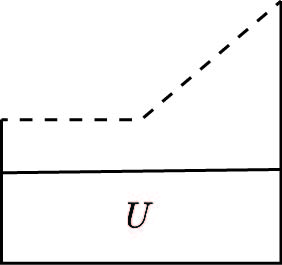} 
 \caption{The spacetime $(M',g)$ from Example 3. Because of the `missing' region (above the dotted line), the only global isometry is the identity map. But there is a non-trivial reflection isometry from the initial segment $U$ to itself.}.
\end{figure}
 
Examples 2 and 3 show the unusual character of de dicto* determinism. Example 2 shows a sense in which the definition is quite strong since it will count as indeterministic some collections of spacetimes that might seem intuitively deterministic. Strong as it is, Example 3 shows a sense in which the definition is also quite weak since it does not imply either of the de re forms of MLE determinism. We see that the agreement of the initial segment $U$ with itself given by $\varphi$ does not extend to a global agreement. So the example does not satisfy the de re forms of MLE determinism. Of course, the agreement of the initial segment $U$ with itself given by $\psi_{|U}$ (the identity map on $U$) does extend to a global agreement via $\psi$ (the identity map on $M'$). Moreover, this global agreement is unique which ensures the example satisfies the de dicto* form of MLE determinism. 

Stepping back, we see that the curious state of affairs obtains because there are non-trivial symmetries of the initial segment $U$ but no such global symmetries. An analogous situation obtains in the collapsing columns example of Wilson (1993). At least, that is the case if the columns collapse in a sufficiently asymmetric way (see Dewar [2025], footnote 28). This suggests that the de dicto* form of MLE is not strong enough by the lights of certain haecceitist positions. In the next section, we will explore the possibility of stronger MLE definitions of determinism that have the effect of ruling out non-trivial local symmetries like the reflection $\varphi$ in Example 3. Here, we continue to explore the de dicto* form and its relationship to other definitions of determinism. 

Recall that Example 1 is the collection $\mathscr{C}=\{(M,g)\}$ where $(M,g)$ is the $t<0$ portion of Minkowski spacetime in standard $(t,x)$ coordinates. Because $\mathscr{C}$ is a singleton collection, it is immediate that it is de dicto deterministic. But it is not de dicto* deterministic since there is a non-trivial global reflection isometry $\psi: M \rightarrow M$ given by $\psi(t,x)=(t, -x)$. We now see that Examples 1, 2, and 3 give us the following:\\

\noindent {\bf Proposition 4:} If a collection $\mathscr{C} \subseteq \mathscr{H}$ is de dicto* deterministic, then it is de dicto deterministic; the other direction does not hold. Moreover, neither de dicto* determinism nor de re* determinism implies the other condition. \\

The entire situation so far is captured in the diagram presented here. Arrows correspond to implication relations. If two conditions in the diagram are not connected by an arrow (or series of arrows), then the corresponding implication relation does not hold. We emphasize that these implication relations hold within (standard) general relativity but not generally (e.g. Leibnizian spacetime is de re but not de re* deterministic).

\[
\begin{array}{ccc}
\text{(de re*)}&&\text{(de dicto*)}\\
\Updownarrow&& \Downarrow\\
\text{(de re)}&\Rightarrow&\text{(de dicto)}\\
\end{array}
\]

\vspace{.5cm}

Why might one be interested in de dicto* determinism? Belot ([1995]) has argued the de dicto, de re, and de re* forms of MLE determinism all fail to be strong enough to count all instances of indeterminism as such. Near the end of the paper, an example is presented. Belot claims the example is deterministic according to de re* determinism (and hence the de re and de dicto forms as well) and yet it is indeterministic by the lights of certain haecceitist positions (for related examples, see Melia [1999]; Pooley [2021]). Here is Belot ([1995], p. 193): `In this example, $W$ is a world with spacetime points and Newtonian
spacetime structure. It initially contains a single $\alpha$ particle. The laws of nature decree that thirteen years later, at $t = 0$, the $\alpha$ particle decays into continuum many $\beta$ particles; arranged so that at time $t$, the $\beta$ particles form a spherical shell of radius $t$; with each $\beta$ particle moving away from the center of the sphere along its radius.'

A global rotational isomorphism of the model $W$ (that preserves both the metric and matter content) takes the worldline of one particle ($\beta_1$) into the worldline of another ($\beta_2$). The presence of such a symmetry supposedly allows one to make sense of the idea that `$\beta_1$ {\em could have been} $\beta_2$' which signals the presence of a type of indeterminism (Belot [1995], p. 194). 

One can object to this line of reasoning (Brighouse [1997]). Indeed, it has been argued that the example is ambiguous as it stands (Halvorson et al. [unpublished]): one natural way of making it precise renders it de re* deterministic while another way -- a less natural approach using names for the $\beta$ particles -- renders it de re* indeterministic. Here, we wish to highlight that even if one holds that both (i) the example is de re* deterministic and also (ii) `$\beta_1$ {\em could have been} $\beta_2$' is a meaningful statement, then there exist other versions of MLE determinism -- e.g. de dicto* -- that have the effect of ruling out this type of symmetric example within the context of general relativity. These other definitions are of independent interest since they will allow us to show that general relativity is deterministic in ways that have not been previously appreciated. 

Consider any collection $\mathscr{C} \subseteq \mathscr{H}$ which counts as de dicto* deterministic. It follows that for any  $(M,g), (M',g') \in \mathscr{C}$, if there is an isometry $\psi: M \rightarrow M'$, it is unique. So every isometry from a spacetime $(M,g)$ to itself must be the identity map; all non-trivial (e.g. rotational) symmetries that take one world line into another are ruled out. This illustrates the way in which de dicto* determinism differs from de re* determinism (which permits non-trivial symmetries).

We close this section with a question that concerns a connection between de dicto* determinism and the theorem of Choquet-Bruhat and Geroch ([1969]). Note that because of Example 2 (the singleton collection of Minkowski spacetime), there is no analogue to Corollary 2 for de dicto* determinism: some collections $\mathscr{C} \subseteq \mathscr{V}^+$ of four-dimensional, inextendible, globally hyperbolic, vacuum solutions to Einstein's equation fail to be de dicto* deterministic. So which sub-collections of $\mathscr{V}^+$ do count as de dicto* deterministic? Consider the following asymmetry condition (Barrett et al. [2023]): \\

\noindent {\bf Definition 6:} A collection $\mathscr{C} \subseteq \mathscr{U}$ is giraffe if, for any $(M,g), (M',g') \in \mathscr{C}$, if there are isometries $\varphi, \psi: M \rightarrow M'$, then $\varphi=\psi$.\\

One can verify that any collection $\mathscr{C} \subseteq \mathscr{U}$ that is giraffe must also be rigid. The other direction does not hold since the collection $\mathscr{U}$ is rigid from Proposition 3 but fails to be giraffe since it contains Minkowski spacetime. Moreover, it follows easily that any collection $\mathscr{C} \subseteq \mathscr{H}$ is de dicto* deterministic if and only if it is both de dicto deterministic and giraffe. To sum up: the isometry uniqueness clause which is added to de dicto determinism to obtain de dicto* determinism has the effect of requiring collections of spacetimes to be giraffe. 

In the same way that the rigidity condition is a type of dual to de re determinism, we see that the giraffe condition is a type of dual to de dicto determinism: the latter condition requires the existence of an isometry $\psi: M \rightarrow M'$ while the former requires its uniqueness. Because any collection $\mathscr{C} \subseteq \mathscr{H}$ that is both giraffe and de dicto deterministic must be de dicto* deterministic, Proposition 2 implies the following: \\

\noindent {\bf Corollary 3:} Any giraffe collection $\mathscr{C} \subseteq \mathscr{V}^+$ is de dicto* deterministic.\\

One naturally wonders: is there a non-empty giraffe collection $\mathscr{C} \subseteq \mathscr{V}^+$? Presumably there is although we are not aware of any result showing this. It is easy to come up with an example that is a four-dimensional, globally hyperbolic, vacuum solution with trivial global symmetries by removing certain regions from Minkowski spacetime. But such an example fails to be inextendible. However, the collection of all spacetimes with trivial global symmetries (i.e. only the identity map as a symmetry) is a giraffe collection and there is reason to think that its spacetimes are, in some sense, `generic' among the collection $\mathscr{U}$ of all spacetimes (see Mounoud [2015], Theorem 1). Perhaps there is a sense in which such spacetimes are generic among the collection $\mathscr{V}^+$ as well.

\vspace{.3cm}

\begin{center}
\large \textbf{5. De Dicto/De Re + Local Uniqueness}
\end{center}

\vspace{.3cm}

\noindent  There is a sense in which de dicto* and de re* forms of MLE determinism, strong though they are, could still be considered too weak by the lights of certain haecceitist positions. The de dicto* form is considered not strong enough since the classical analogue counts the collapsing columns example as deterministic. That example is counted as indeterministic under the classical analogue of the de re* form. But by the lights of certain (in our view, misguided) haecceitist positions, the de re* definition has its own problems. Belot has argued that it is also not strong enough in the sense that it wrongly counts the particle decay example as deterministic. 

In both examples, symmetries are present; local symmetries are present in the initial segment of the collapsing columns example and global symmetries are present in the particle decay example. This naturally suggests a way to strengthen each definition: rule out all symmetries. We shall do this by adding a `local uniqueness' clause to both the de dicto and de re forms of MLE determinism: even parts of worlds (including initial segments) must agree in only one way. 

For any spacetimes $(M,g), (M',g') \in \mathscr{U}$, let us say that a map $\psi: M \rightarrow M'$ is a (not necessarily proper) `isometric embedding' if $\psi$ is an isometry from $M$ to $\psi[M]$. Consider the following local asymmetry condition (Manchak and Barrett [forthcoming]): \\

\noindent {\bf Definition 7:} A collection $\mathscr{C} \subseteq \mathscr{U}$ is Heraclitus if, for any $(M,g), (M',g') \in \mathscr{C}$ and any open set $O \subseteq M$, if there are isometric embeddings $\varphi, \psi: O \rightarrow M'$, then $\varphi=\psi$.\\

The Heraclitus condition ensures that even local isometries between spacetimes must be unique. Let $\mathscr{C} \subseteq \mathscr{U}$ be any Heraclitus collection and consider any $(M,g), (M',g') \in \mathscr{C}$. For any initial segment $U \subset M$, we find there is at most one initial segment $U' \subset M'$ isometric to $U$. Moreover, if there is an isometry relating $U$ and $U'$, it is unique. One can verify that any collection that is Heraclitus must also be giraffe (and thus rigid). The other direction does not hold since the collection $\mathscr{C}=\{(M',g)\}$ from Example 3 is giraffe but fails to be Heraclitus. Because the Heraclitus asymmetry condition is the natural local strengthening of the giraffe asymmetry condition, we immediately see the natural way to strengthen the de dicto* and de re* forms of determinism -- at least within the context of general relativity. We have the following:  \\

\noindent {\bf Definition 8:} A collection $\mathscr{C} \subseteq \mathscr{H}$ is de dicto** deterministic if it is de dicto and Heraclitus. \\

\noindent {\bf Definition 9:} A collection $\mathscr{C} \subseteq \mathscr{H}$ is  is de re** deterministic if it is de re and Heraclitus.\\

Because the Heraclitus condition in each of the definitions ensures that even local isometries between spacetimes must be unique, it serves to rule out even the local version of a `$\beta_1$ {\em could have been} $\beta_2$' argument. Because of Examples 2 and 3 and the way we have set things up, we have the following:\\

\noindent {\bf Proposition 5:} If a collection $\mathscr{C} \subseteq \mathscr{H}$ is de dicto** deterministic, then it is de dicto* deterministic (and thus de dicto deterministic); the other direction does not hold. If a collection $\mathscr{C} \subseteq \mathscr{H}$ is de re** deterministic, then it is de re* deterministic (and thus de re deterministic); the other direction does not hold.\\

Curiously, we find that de dicto** determinism and de re** determinism are equivalent. Consider the following: \\

\noindent {\bf Proposition 6:} A collection $\mathscr{C} \subseteq \mathscr{H}$ is de dicto** deterministic if and only if it is de re** deterministic.\\

\noindent {\bf Proof:} One direction is trivial. Suppose a collection $\mathscr{C} \subseteq \mathscr{H}$ is de dicto** deterministic. So $\mathscr{C}$ is Heraclitus. We need only show that it is also de re deterministic. Let $(M,g)$ and $(M',g')$ be any spacetimes in $\mathscr{C}$ and let $U \subset M$ and $U' \subset M'$ be any initial segments. Suppose there is an isometry $\varphi: U \rightarrow U'$. Since $\mathscr{C}$ is de dicto** deterministic, it is de dicto deterministic. So there is an isometry $\psi: M \rightarrow M'$. We find that $\psi_{|U}$ and $\varphi$ are both isometric embeddings of the open set $U$ into $M'$. Because $\mathscr{C}$ is Heraclitus, we have $\psi_{|U}=\varphi$. Thus $\mathscr{C}$ is de re deterministic and we are done. \qed \\

The entire situation is captured in the diagram presented here. As before, arrows correspond to implication relations. If two conditions in the diagram are not connected by an arrow (or series of arrows), then the corresponding implication relation does not hold. We emphasize that, although these implication relations hold within general relativity, there are open questions here about what the situation looks outside of this context. (All down arrows and right arrows depicted in the diagram hold in general. But as we have seen, Leibnizian spacetime satisfies the classical analogue to de re determinism but not de re* determinism. One wonders about the situation regarding de dicto** and de re** determinism within the context of various classical spacetime theories. Do classical Heraclitus spacetimes even exist?)  

\[
\begin{array}{ccc}
\text{(de re**)}&\Leftrightarrow&\text{(de dicto**)}\\
\Downarrow&& \Downarrow\\
\text{(de re*)}&&\text{(de dicto*)}\\
\Updownarrow&& \Downarrow\\
\text{(de re)}&\Rightarrow&\text{(de dicto)}\\
\end{array}
\]

\vspace{.5cm}

Despite the strength of de dicto**/de re** determinism, these definitions are still weak enough to be satisfied by non-empty collections of spacetimes. Consider the following:\\

\noindent {\bf Example 4:} Let $(M,g)$ be the two-dimensional spacetime constructed in Manchak and Barrett ([forthcoming], pp. 11-14). This spacetime has the `Heraclitus' property of spacetime which requires that, for any open regions $U, U' \subseteq M$, if there is an isometry $\psi: U \rightarrow U'$, then $U=U'$ and $\psi$ is the identity map. The example is conformally flat with a manifold diffeomorphic to $\mathbb{R}^2$. The Heraclitus property obtains because of a careful choice of conformal factor which ensures that no distinct points in $M$ have all of the same scalar curvature values. Let $(M',g)$ be any globally hyperbolic region of $(M,g)$. (We know that arbitrarily small such regions exist around any point in any spacetime.) It follows easily that $(M',g)$ inherits the Heraclitus property of spacetime from $(M,g)$. So $\mathscr{C}=\{(M',g)\}$ must be a Heraclitus collection. But $\mathscr{C}$ is also de dicto deterministic. This follows easily since there is only one spacetime in the collection. Since $\mathscr{C}$ is both Heraclitus and de dicto deterministic, it is de dicto** deterministic. Thus, by Proposition 6, it is de re** deterministic as well.\\

We now consider a question analogous to one posed at the end of the previous section. It concerns a connection between de dicto**/de re** determinism and the theorem of Choquet-Bruhat and Geroch ([1969]). Recall that $\mathscr{V}^+$ is the collection of four-dimensional, inextendible, vacuum, globally hyperbolic spacetimes. Which sub-collections of $\mathscr{V}^+$ do count as de dicto**/de re** deterministic? Because any collection $\mathscr{C} \subseteq \mathscr{H}$ that is both Heraclitus and de dicto deterministic must be de dicto** deterministic, Propositions 2 and 6 imply the following:\\

\noindent {\bf Corollary 4:} Any Heraclitus collection $\mathscr{C} \subseteq \mathscr{V}^+$ is de dicto** deterministic and thus de re** deterministic.\\

As before, one naturally wonders: is there a non-empty Heraclitus collection $\mathscr{C} \subseteq \mathscr{V}^+$? As with the giraffe case, presumably there is. The collection of all spacetimes with trivial local symmetries is a Heraclitus collection and there is reason to think that its spacetimes are, in some sense, `generic' among the collection $\mathscr{U}$ of all spacetimes (see Sunada [1985], Proposition 1). 

We close by drawing attention to a strong connection between de dicto**/ de re** determinism and another (more metaphysical) way of capturing MLE determinism which requires an `intrinsic' agreement of possible worlds. This form of determinism goes by many names. It is unfortunately called `de re' by Hawthorne ([2006]). It is also called `Det1' by Pooley ([2021]) and `full' by Teitel ([2019]) and Halvorson et al. ([unpublished]). To avoid confusion, let us follow the latter usage: a collection of possible worlds satisfies `full determinism' if, for any worlds in the collection that `intrinsically' agree on some initial segment, the worlds `intrinsically' agree in their entirety. 

It would seem that in order to make sense of `intrinsic' agreement, one must restrict attention to theories in which objects are somehow named, e.g. the $\beta$ particles in Belot's particle decay example. This is because full determinism requires that the theory specify which object in its domain has what future properties (see Halvorson et al. ([unpublished]) for an extended discussion of this matter). Due to the possibility of symmetries, collections of general relativistic spacetimes will not have such names in general. (Here, we are following standard practice and assuming that names (whatever they may be -- properties, constants, etc) must be preserved by model isomorphisms. So if all points have names, then ipso facto the model is not symmetric.)

But we emphasize that collections of general relativistic spacetimes that satisfy de dicto**/ de re** determinism are necessarily Heraclitus. This means that each point in each spacetime is effectively named since isometries (even local isometries) among Heraclitus spacetimes are unique. In Example 4, one can verify that each spacetime point can be distinguished from any other via its unique invariant curvature properties. So we see that a collection of spacetimes that is sufficiently asymmetric will have sufficiently many names to make sense of the `intrinsic' agreement of possible worlds. It follows that a collection of general relativistic spacetimes that satisfies de dicto**/de re** determinism will also satisfy full determinism (Halvorson et al. [2025]). Since we know that there are collections of general relativistic spacetimes that satisfy de dicto**/de re** determinism (e.g. the one given in Example 4), we know that there are collections of general relativistic spacetimes that satisfy full determinism. 

\vspace{.3cm}

\begin{center}
\large \textbf{6. Conclusion}
\end{center}

\vspace{.3cm}

\noindent  We have considered several precise formulations of MLE determinism idea: if worlds agree on some initial segment, then the worlds agree entirely. We emphasize that we have not tried to capture, once and for all, a single notion of `determinism' within general relativity. In a pluralistic way, we have simply mapped out which collections of general relativistic spacetimes are deterministic relative to which definitions. We have also clarified the relationships between the definitions and posed some open questions. Here is a summary of the situation.

The de dicto and de re formulations of MLE determinism have long played a central role in discussions of the hole argument (Butterfield [1989]; Melia [1999]; Pooley [2021]). Within that context, de re* determinism was first introduced by Belot ([1995]). This definition introduces a uniqueness clause into the consequent of the de re form of MLE determinism: worlds must agree in only one way. Within this context, this uniqueness clause amounts to the asymmetry condition of rigidity (Geroch [1969]) which has recently become a focus of attention in the hole argument literature (Menon and Read [2023]; Dewar [2025]; Halvorson and Manchak [2025]). We have seen how a generalized version of de re* determinism is a sufficient condition on spacetime theories to ensure that the identity map is not an isomorphism between models related by a non-trivial hole diffeomorphism (cf. Weatherall [2018]). The same cannot be said for generalized versions of de re or de dicto determinism. Moreover, unlike the de dicto and de re forms of determinism, the de re* form is strong enough to capture the intuition of Earman ([1977]) that Leibnizian spacetime is indeterministic. Even so, by the lights of certain haecceitist positions, de re* determinism is still too weak in the sense that Belot's particle decay example is wrongly counted as deterministic under this form. 

One can object to the significance of the particle decay example (Brighouse [1997]). Indeed, it has been argued that one cannot hold both of the following: (i) the example counts as de re* deterministic and (ii) it makes sense to say that `$\beta_1$ {\em could have been} $\beta_2$' (Halvorson et al. [unpublished]). But in this paper, we have explored an altogether different type of response. We have looked for even stronger versions of MLE determinism to rule out such examples. We first considered de dicto* determinism which is the analogue to re de* determinism. This definition introduces a uniqueness clause which amounts to the giraffe asymmetry condition (Barrett et al. [2023]). We found that de dicto* determinism is neither stronger nor weaker than de re* determinism but that it is strong enough to rule out the globally symmetric problem cases (e.g. the particle decay example) central to the hole argument literature. Even so, it still may not be strong enough by the lights of certain haecceitist positions: some locally symmetric examples that seem to be indeterministic (e.g. the collapsing columns) are not counted as such under the de dicto* form.  

We then introduced a local uniqueness clause into both the de dicto and de re forms of determinism: even parts of worlds (including initial segments) must agree in only one way. This local uniqueness clause amounts to the Heraclitus asymmetry condition (Manchak and Barrett [forthcoming]). The resulting definitions of de dicto** and de re** determinism turn out to be equivalent and also seem to be sufficiently strong to rule out, not only the globally symmetric problem cases, but locally symmetric variants as well. We have shown that the satisfaction of de dicto**/re re** determinism by a collection of general relativistic spacetimes implies that each point in each spacetime is effectively named since isometries (even local isometries) among Heraclitus spacetimes are unique. This allows one to make sense of `intrinsic' agreement of possible worlds which is required by `full' determinism (Hawthorne [2006]; Teitel [2019]; Pooley [2021]). It follows that any collection of general relativistic spacetimes that satisfies de dicto**/de re** determinism will also satisfy full determinism. We have shown that de dicto**/de re** determinism, strong as it is, is weak enough to be satisfied by some non-empty collections of general relativistic spacetimes. It is an open question whether there are significant swathes of general relativistic spacetimes of physical interest that are de dicto**/de re** deterministic. Perhaps there are given that the Heraclitus condition seems to be generically satisfied (see Sunada [1985], Proposition 1). 

\vspace{.5cm}

\begin{center}
\large \textbf{Acknowledgements}
\end{center}

\vspace{.3cm}

\noindent Thanks to Jeremy Butterfield, Neil Dewar, David Malament, Oliver Pooley, and two anonymous referees for helpful comments on previous drafts. JOW: This material is based upon work supported by the National Science Foundation under Grant No. 2419967.

\vspace{1.2cm}

\begin{flushright}
JB Manchak\\
{\em Department of Logic and Philosophy of Science\\
University of California, Irvine\\
Irvine, CA, USA\\
jmanchak@uci.edu}\\

\vspace{1cm}

TW Barrett\\
{\em Department of Philosophy\\
University of California, Santa Barbara\\
Santa Barbara, CA, USA\\
tbarrett@philosophy.ucsb.edu}\\

\vspace{1cm}

HP Halvorson\\
{\em Department of Philosophy\\
Princeton University\\
Princeton, NJ, USA\\
hhalvors@princeton.edu}\\

\vspace{1cm}

JO Weatherall\\
{\em Department of Logic and Philosophy of Science\\
University of California, Irvine\\
Irvine, CA, USA\\
weatherj@uci.edu}\\
\end{flushright}

\vspace{.3cm}

\begin{center}
\large \textbf{References}
\end{center}

\noindent \hangindent=.5cm Barrett, T., Manchak, J., and Weatherall, J. [2023]: `On Automorphism Criteria for Comparing Amounts of Mathematical Structure', {\em Synthese}, {\bf 201}, pp. 1-14. 

\noindent \hangindent=.5cm Belot, G. [1995]: `New Work for Counterpart Theorists: Determinism', {\em The British Journal for the Philosophy of Science}, {\bf 46}, pp. 185-195.

\noindent \hangindent=.5cm Brighouse, C. [1997]: `Determinism and Modality', {\em The British Journal for the Philosophy of Science}, {\bf 48}, pp. 465-481.

\noindent \hangindent=.5cm Butterfield, J. [1989]: `The Hole Truth', {\em The British Journal for the Philosophy of Science}, {\bf 40}, pp. 1-28.

\noindent \hangindent=.5cm Choquet-Bruhat, Y. and Geroch, R. [1969]: `Global Aspects of the Cauchy Problem in General Relativity', {\em Communications in Mathematical Physics}, {\bf 14}, pp. 329-335. 

\noindent \hangindent=.5cm Cudek, F. [forthcoming]: `Counterparts, Determinism, and the Hole Argument', {\em The British Journal for the Philosophy of Science}.

\noindent \hangindent=.5cm Curie, P. [1894]: `Sur la Sym\'{e}trie dans les Ph\`{e}nom\'{e}nes Physique, Sym\'{e}trie d'un Champ \'{E}lectrique et d'un Champ Magn\'{e}tique', {\em Journal de Physique Th\'{e}orique et Appliqu\'{e}e}, {\bf 3}, pp. 393-415.

\noindent \hangindent=.5cm Dewar, N. [2016]: {\em Symmetries in Physics, Metaphysics, and Logic}, PhD thesis, University of Oxford.

\noindent \hangindent=.5cm Dewar, N. [2025]: `The Hole Argument and Determinism(s)', {Philosophy of Physics}, {\bf 3}, 10.

\noindent \hangindent=.5cm Doboszewski, J. [2019]: `Relativistic Spacetimes and Definitions of Determinism', {\em European Journal for the Philosophy of Science}, {\bf 9}, 24. 

\noindent \hangindent=.5cm Earman, J. [1977]: `Leibnizian Space-Times and Leibnizian Algebras', in R. Butts and J. Hintikka (eds.), {\em Historical and Philosophical Dimensions of Logic, Methodology and Philosophy of Science}, Reidel Publishing, pp. 93-112.

\noindent \hangindent=.5cm Earman, J. [1986]: {\em A Primer on Determinism}, Reidel Publishing. 

\noindent \hangindent=.5cm Earman, J. [1989]: {\em World Enough and Space-Time}, MIT Press.

\noindent \hangindent=.5cm Earman, J. [2007]: `Aspects of Determinism in Modern Physics', in J. Butterfield and J. Earman (eds.), {\em Philosophy of Physics}, Elsevier, pp. 1369-1434.

\noindent \hangindent=.5cm Earman, J. and Norton, J. [1987]: `What Price Spacetime Substantivalism? The Hole Story', {\em The British Journal for the Philosophy of Science}, {\bf 38}, pp. 515-525.

\noindent \hangindent=.5cm Geroch, R. [1969]: `Limits of Spacetimes', {\em Communications in Mathematical Physics}, {\bf 13}, pp. 180-193.

\noindent \hangindent=.5cm Halvorson, H. and Manchak, J. [2025]: `Closing the Hole Argument', {\em The British Journal
for the Philosophy of Science}, {\bf 76}, pp. 295-318. 

\noindent \hangindent=.5cm Halvorson, H. Manchak, J. and Weatherall, J. [unpublished]: `Defining Determinism', available at $<$https://arxiv.org/pdf/2503.05681$>$.

\noindent \hangindent=.5cm Hawthorne, J. [2006]: {\em Metaphysical Essays}, Oxford University Press.

\noindent \hangindent=.5cm Landsman, K. [2023]: `Reopening the Hole Argument', {\em Philosophy of Physics}, {\bf 1}, 6. 

\noindent \hangindent=.5cm Lewis, D. [1983]: `New Work for a Theory of Universals', {\em Australasian Journal of Philosophy}, {\bf 61}, pp. 343-377.

\noindent \hangindent=.5cm Manchak, J. [2016]: `Is the Universe As Large As It Can Be?', {\em Erkenntnis}, {\bf 81}, pp. 1341-1344.

\noindent \hangindent=.5cm Manchak, J. [2017]: `On the Inextendibility of Spacetime', {\em Philosophy of Science}, {\bf 84}, pp. 1215-1225.

\noindent \hangindent=.5cm Manchak, J. [2020]: {\em Global Spacetime Structure}, Cambridge University Press. 

\noindent \hangindent=.5cm Manchak, J. [2023]:  `On the (In?)Stability of Spacetime Inextendibility', {\em Philosophy of Science}, {\bf 90}, pp. 1331-1341.

\noindent \hangindent=.5cm Manchak, J. [unpublished]: {\em Is the Universe As Large As It Can Be? A Study of Spacetime Possibility}, available at $<$https://philsci-archive.pitt.edu/25330/1/isuasaicbasisp.pdf$>$.

\noindent \hangindent=.5cm Manchak, J. and Barrett, T. [forthcoming]: `A Hierarchy of Spacetime Symmetries: Holes to Heraclitus', {\em The British Journal for the Philosophy of Science}. 

\noindent \hangindent=.5cm Melia, J. [1999]: `Holes, Haecceitism and Two Conceptions of Determinism', {\em The British Journal for the Philosophy of Science}, {\bf 50}, pp. 639-664.

\noindent \hangindent=.5cm Menon, T. and Read, J. [2024]: `Some Remarks on Recent Formalist Responses to the Hole Argument', {\em Foundations of Physics}, {\bf 54}, 6.

\noindent \hangindent=.5cm Montague, R. [1974]: `Deterministic Theories', in R. Thomason (ed.), {\em Formal Philosophy}, Yale University Press, pp. 303-359.

\noindent \hangindent=.5cm Mounoud, P. [2015]: `Metrics Without Isometries Are Generic', {\em Monatshefte f\'ur Mathematik}, {\bf 176}, pp. 603-606. 

\noindent \hangindent=.5cm Pooley, O. [2021]: `The Hole Argument', in E. Knox and A. Wilson (eds.), {\em The Routledge Companion to Philosophy of Physics}, Routledge, pp. 145-159.

\noindent \hangindent=.5cm Roberts, B. [2016]: `Curie's Hazard: From Electromagnetism to Symmetry Violation', {\em Erkenntnis}, {\bf 81}, pp. 1011-1029.

\noindent \hangindent=.5cm Roberts, B. [2020]: `Regarding `Leibniz Equivalence', {Foundations of Physics}, {\bf 50}, pp. 250-269. 

\noindent \hangindent=.5cm Smeenk, C. and W\"uthrich, C. [2021]: `Determinism and General Relativity', {\em Philosophy of Science}, {\bf 88}, pp. 638-664. 

\noindent \hangindent=.5cm Stein, H. [1977]: `Some Philosophical Prehistory of General Relativity',  in J. Earman, C. Glymour, and J. Stachel (eds.), {\em Foundations of Space-Time Theories, Minnesota Studies in the Philosophy of Science (volume 8)}, University of Minnesota Press, pp. 3-49.

\noindent \hangindent=.5cm Sunada, T. [1985]: `Riemannian Coverings and Isospectral Manifolds', {\em Annals of Mathematics}, {\bf 121}, pp. 169-186. 

\noindent \hangindent=.5cm Teitel, T. [2019]: `Holes in Spacetime: Some Neglected Essentials', {\em Journal of Philosophy}, {\bf 116}, pp. 353-389.

\noindent \hangindent=.5cm Wald, R. [1984]: {\em General Relativity}, University of Chicago Press.

\noindent \hangindent=.5cm Weatherall, J. [2018]: `Regarding the Hole Argument', {\em The British Journal for the Philosophy of Science}, {\bf 69}, pp. 329-350.

\noindent \hangindent=.5cm Weatherall, J. [2020]: `Some Philosophical Prehistory of the (Earman-Norton) Hole Argument', {\em Studies in History and Philosophy of Modern Physics}, {\bf 70}, pp. 79-87.

\noindent \hangindent=.5cm Wilson, M. [1993]: `There's a Hole and a Bucket, Dear Leibniz', {\em Midwest Studies in Philosophy}, {\bf 18}, pp. 202-241.

\end{document}